\documentclass[
showpacs,
floatfix,
aps,
prb,
twocolumn,
superscriptaddress,
amssymb,
]{revtex4-1}

\usepackage{times}
\usepackage{amssymb}
\usepackage{latexsym}
\usepackage[dvips]{graphicx}
\usepackage{amsmath}
\usepackage{graphicx}
\usepackage{dcolumn}
\usepackage{amsfonts}
\usepackage{bm}
\usepackage{epsfig}

\newcommand{\be}{\begin{equation}}
\newcommand{\ee}{\end{equation}}
\newcommand{\bea}{\begin{eqnarray}}
\newcommand{\eea}{\end{eqnarray}}

\def\xt{\mathcal{X}_{2}}

\def\m{m^\star}

\def\oc{\omega_{\mbox{\scriptsize {c}}}}
\def\op{\omega_{\mbox{\scriptsize {p}}}}

\def\tem{\tau_{\it em}}

\newcommand{\req}[1]{Eq.\,(\ref{#1})}

\newcommand{\rfig}[1]{Fig.\,\ref{#1}}

\newcommand{\rref}[1]{Ref.\,\onlinecite{#1}}

\begin{document}
\title{
Magnetoplasmon resonance in 2D electron system driven into a zero-resistance state
}
\author{A.\,T. Hatke}
\affiliation{School of Physics and Astronomy, University of Minnesota, Minneapolis, Minnesota 55455, USA}

\author{M.\,A. Zudov}
\email[Corresponding author: ]{zudov@physics.umn.edu}
\affiliation{School of Physics and Astronomy, University of Minnesota, Minneapolis, Minnesota 55455, USA}

\author{J.\,D. Watson}
\affiliation{Department of Physics, Purdue University, West Lafayette, Indiana 47907, USA}
\affiliation{Birck Nanotechnology Center, School of Materials Engineering and School of Electrical and Computer Engineering, Purdue University, West Lafayette, Indiana 47907, USA}

\author{M.\,J. Manfra}
\affiliation{Department of Physics, Purdue University, West Lafayette, Indiana 47907, USA}
\affiliation{Birck Nanotechnology Center, School of Materials Engineering and School of Electrical and Computer Engineering, Purdue University, West Lafayette, Indiana 47907, USA}

\received{January 12, 2012; accepted for publication March 21, 2012 }

\begin{abstract}
We report on a remarkably strong, and a rather sharp, photoresistance peak originating from a dimensional magnetoplasmon resonance (MPR) in a high mobility GaAs/AlGaAs quantum well driven by microwave radiation into a zero-resistance state (ZRS).
The analysis of the MPR signal reveals a negative background, providing experimental evidence for the concept of absolute negative resistance associated with the ZRS.
When the system is further subject to a dc field, the maxima of microwave-induced resistance oscillations decay away and the system reveals a state with close-to-zero differential resistance. 
The MPR peak, on the other hand, remains essentially unchanged, indicating surprisingly robust Ohmic behavior under the MPR conditions.

\end{abstract}
\pacs{73.43.Qt, 73.63.Hs, 73.40.-c}
\maketitle

Recent low-field magnetotransport experiments in very high mobility two-dimensional electron systems (2DES) revealed a variety of remarkable phenomena,\cite{zudov:2001a,zudov:2001b,yang:2002,mani:2002,zudov:2003,yang:2003,kukushkin:2004,bykov:2007,khodas:2010}
which include microwave-induced resistance oscillations (MIRO).\citep{zudov:2001a,studenikin:2005,studenikin:2007,hatke:2009a,hatke:2011e} 
MIRO originate from either the {\em displacement} mechanism,\citep{ryzhii:1970,durst:2003,lei:2003,vavilov:2004,dmitriev:2009b} stepping from the modification of impurity scattering by microwaves, or from the {\em inelastic} mechanism,\citep{dmitriev:2005,dmitriev:2009b} owing to the radiation-induced non-equilibrium distribution of electrons.
In either case, MIRO can be described by a radiation-induced correction (photoresistivity) of the form
\be
\delta\rho_\omega \propto - 
\sin(2\pi\omega/\oc)\,,
\label{eq.miro}
\ee
where
$\oc=eB/\m$ is the cyclotron frequency, $\m$ is the electron effective mass, and $\omega=2\pi f$ is the microwave frequency.
The negative photoresistance at the MIRO minima can approach (but cannot exceed) the dark resistivity, by absolute value, giving rise to zero-resistance states (ZRS).\citep{mani:2002,zudov:2003,willett:2004,smet:2005,zudov:2006a,zudov:2006b,bykov:2006,dorozhkin:2011} 
It was predicted theoretically\citep{andreev:2003} that ZRS emerge as a result of an instability of the underlying negative resistance.

In addition to MIRO, microwave photoresistance can also reveal magnetoplasmon resonance (MPR).\citep{vasiliadou:1993,zudov:2001a,kukushkin:2006b,yang:2006,dorozhkin:2007,tung:2009,andreev:2011}
The dispersion of 2D plasmons in the long-wavelength limit was calculated by \textcite{stern:1967} 
\be
\op^2(q)=\frac{e^2n_e}{2\varepsilon_0\bar\varepsilon m^*}q\,,
\label{eq.p}
\ee
where $\varepsilon_0$ is the permittivity of vacuum, $\bar\varepsilon$ is the effective dielectric constant of the surroundings,\citep{note:5} and $n_e$ is the density of 2D electrons. 
In a laterally confined 2DES, such as a long Hall bar of width $w$, fundamental mode of standing plasmon oscillations have a wave number $q_0 = \pi/w$.
Upon application of a perpendicular magnetic field $B$, the plasmon mode hybridizes with the cyclotron resonance \citep{chaplik:1972} and the dispersion of a combined (magnetoplasmon) mode is given by
\be
\omega^2=\oc^2+\op^2\,.
\label{eq.mp}
\ee
In contrast to MIRO, there exists no theory of MPR photoresistance.
However, it is believed that radiation absorption translates to electron heating which, in turn, causes a (usually positive\citep{note:6}) 
resistivity change.\citep{vasiliadou:1993,kukushkin:2006b}

While both MPR and MIRO were realized simultaneously in several experiments,\citep{zudov:2001a,yang:2006,yuan:2006,dorozhkin:2007,tung:2009} the MPR peak remained much smaller than both MIRO and the dark resistivity.
Moreover, even in studies using ultra-high mobility 2DES, the MPR peak remained weak but was broad enough to completely destroy ZRS,\citep{yang:2006} and, as a result, their interplay could not be explored.
On the other hand, it is interesting to see if the MPR can be used to study ZRS and, e.g., to obtain information on the underlying absolute negative resistance predicted almost a decade ago.\citep{andreev:2003}
Furthermore, there exist no studies of the MPR in strong dc electric fields, which were successfully used to get insight into other low field phenomena.\citep{zhang:2007c,hatke:2008a,hatke:2008b,hatke:2011c}

In this Rapid Communication we report on microwave photoresistivity measurements in a high mobility GaAs/AlGaAs quantum well.
In addition to MIRO and ZRS, our experiment reveals a remarkably strong and sharp photoresistance peak.
This peak originates from a dimensional MPR and, in contrast to previous studies, its height is comparable to the MIRO amplitude, to the zero-field resistivity, and is several times larger than the dark resistivity.
By tuning the microwave frequency, the MPR and ZRS conditions can be satisfied simultaneously giving rise to a re-entrant ZRS interrupted by the sharp MPR peak.
Lorentzian fit of the MPR peak reveals a negative background, providing strong evidence for the absolute negative resistance associated with ZRS.\citep{andreev:2003}
Upon application of a dc electric field, low-order MIRO maxima quickly decay and the 2DES goes into a state with close-to-zero differential resistance.\citep{hatke:2010a} 
The MPR peak, on the other hand, shows surprisingly little sensitivity to the dc field, both in its magnitude and in its position.
This behavior implies that under the MPR condition, the resistivity remains Ohmic to much larger currents compared to both MIRO and the dark resistivity.

Our sample is a lithographically defined Hall bar (width $w=50$ $\mu$m) fabricated from a 300 \AA-wide GaAs/Al$_{0.24}$Ga$_{0.76}$As quantum well grown by molecular beam epitaxy.
After a brief low-temperature illumination, the density and the mobility were $n_e \approx 2.9 \times 10^{11}$ cm$^{-2}$ and $\mu\simeq 1.3 \times 10^7$ cm$^2$/Vs, respectively. 
Microwave radiation of frequency $f$, generated by a backward wave oscillator, was delivered to the sample via a 1/4 inch (6.35 mm) diameter light pipe. 
The resistivity $\rho_\omega$ and the differential resistivity $r_\omega \equiv dV/dI$ were measured using a low-frequency lock-in technique under continuous microwave irradiation in sweeping magnetic field.

\begin{figure}[t]
\includegraphics{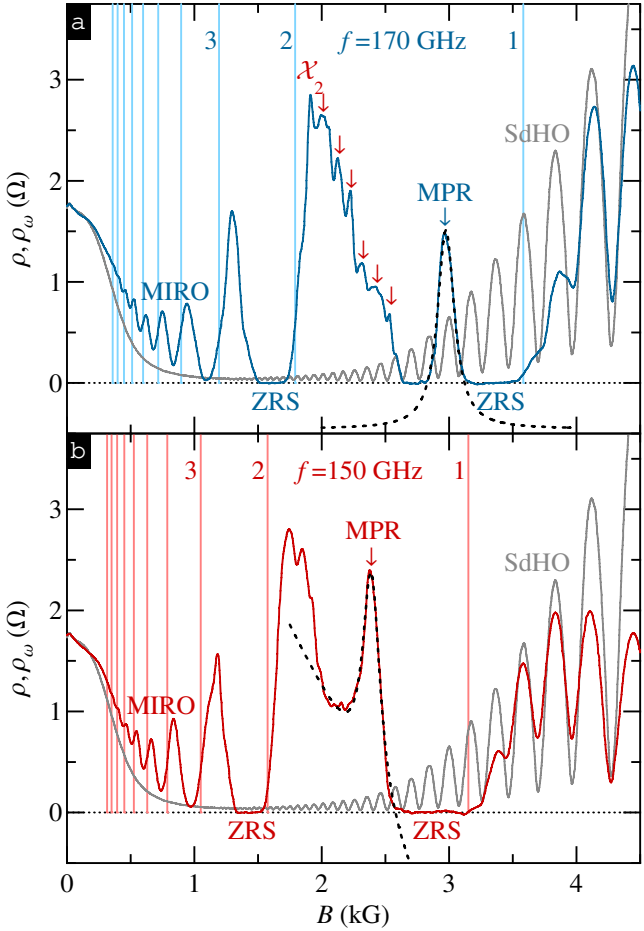}
\vspace{-0.1 in}
\caption{(Color online)
Magnetoresistivity $\rho_\omega(B)$\citep{note:x} (dark curves) under microwave irradiation of frequency (a) $f = 170$ GHz and (b) $f = 150$ GHz at $T = 0.65$ K.
Both panels also show $\rho(B)$ measured without irradiation (light curves).
The dashed curves are fits to the data (see text).
The vertical lines are drawn at the harmonics of the cyclotron resonance, $\omega/\oc=1,2,3,\dots\,$, obtained from the MIRO period.
The MPR and $\xt$ peaks are marked by ``MPR'' and by ``$\xt$'', respectively.
Arrows mark an additional series of peaks.
}
\vspace{-0.15 in}
\label{fig1}
\end{figure}
In \rfig{fig1}(a) and (b) we present resistivity, $\rho_\omega$, (dark curves) as a function of magnetic field $B$ under microwave irradiation of frequency $f = 170$ GHz and $f = 150$ GHz, respectively, measured at $T = 0.65$ K.
For comparison, both panels also show magnetoresistivity, $\rho(B)$, measured without microwave irradiation (light curves).
Without radiation, $\rho(B)$ exhibits a strong negative magnetoresistance effect\citep{dai:2010,hatke:2011b,bockhorn:2011,hatke:2012a} followed by Shubnikov-de Haas oscillations (SdHO) at $B \gtrsim 1.5$ kG. 
Under microwave irradiation, the magnetoresistivity $\rho_\omega(B)$ reveals pronounced MIRO which persisting up to the 10-th order.
Being controlled by $\omega/\oc$ [cf. \req{eq.miro}], MIRO appear near the cyclotron resonance harmonics at both frequencies (cf. vertical lines drawn at $\omega/\oc = 1,2,3,...$).
We further notice that $\rho_\omega(B)$ reveals a series of fast oscillations superimposed on the second MIRO maximum [cf.\,$\downarrow$ in \rfig{fig1}(a)].
At this point we are not certain about the origin of these oscillations\citep{note:3} but the position and the shape of the maximum closest to the second cyclotron resonance harmonic [marked by ``$\xt$'' in \rfig{fig1}(a)] appear consistent with the recently discovered $\xt$ peak,\citep{dai:2010,hatke:2011b,hatke:2011c,dai:2011,hatke:2011f} whose nature, however, is also unknown at this point. 

Further examination of the data reveals that the lower order MIRO minima are developed into ZRS, attesting to the high quality of our 2DES.
Remarkably, the fundamental (first order) ZRS in \rfig{fig1}(a) is interrupted by a very strong and sharp photoresistance peak.
As we show below, this peak (marked by ``MPR'') corresponds to the fundamental mode of the dimensional MPR, which apparently can easily destroy the current domain structure associated with the ZRS.\citep{andreev:2003,dorozhkin:2011}
The height of this MPR peak in our experiment is {\em several times larger} than the dark resistivity. 
This finding contrasts with previous studies,\citep{vasiliadou:1993,zudov:2001a,kukushkin:2006b,yang:2006} where the the MPR photoresistance was only a few percent of the dark resistivity.
While the origin of such a giant response to MPR in our 2DES is not precisely known, it might be qualitatively explained by strong temperature dependence\citep{hatke:2012a} of the dark resistivity in the regime of the giant negative magnetoresistance.\citep{dai:2010,hatke:2011b,bockhorn:2011,hatke:2012a}
Under the MPR condition, this strong temperature dependence translates to a giant resistivity peak owing to electron heating due to resonant absorption of radiation.

We next examine the height, the position, and the width of the MPR peak shown in \rfig{fig1}(a) in more detail.
The peak height, if measured from zero, is about 1.5 $\Omega$, which is comparable to both the zero-field resistivity and the third-order MIRO peak.
However, since the MPR peak is overlapping with the ZRS, which is believed to be characterized by the underlying negative resistance,\citep{andreev:2003,zudov:2006b} the actual height of the peak should be even larger.
To test this prediction, and to obtain other characteristics of the MPR photoresistance, we fit our data with Lorentzian, $\rho_\omega(B) = a + b/[(B-B_0)^2+(\delta B)^2]$, and present the result as a dashed curve in \rfig{fig1}(a).
The fitting procedure reveals the negative background $a \approx - 0.45~\Omega$,\citep{note:10} suggesting that the actual height of the MPR peak in \rfig{fig1}(a) is close to 2.0 $\Omega$, and that the MPR photoresistance can be used to probe the absolute negative resistance associated with the ZRS.\citep{note:8}
We also notice that the half-width of the MPR peak, $\delta B \approx 0.08$ kG $\approx 0.16$ K, is considerably smaller than the radiative decay rate, $\tau_{\rm em}^{-1}=n_e e^{2}/2 \epsilon_{0} \sqrt{\tilde\varepsilon} \m c$, $\sqrt{\tilde\varepsilon}=(\sqrt{12.8}+1)/2 \approx 2.3$,\citep{chiu:1976} which we estimate as $\tem^{-1} \approx 0.74$ K in our 2DES.

With $B_0\approx 3.0$ kG and $m^\star = 0.067\,m_0$, we calculate the plasmon frequency $f_p = \sqrt{\omega^2 - \oc^2}/2\pi \approx 115$ GHz using \req{eq.mp}.
This value is somewhat lower than $\op(q_0)/2\pi \approx 126$ GHz obtained from \req{eq.p}.
We notice that the dispersion given by \req{eq.mp} is generally valid in a quasi-electrostatic approximation where the retardation effects can be ignored.\citep{mikhailov:2004,mikhailov:2005}
According to \rref{mikhailov:2005}, the importance of retardation can be described by the ratio of the plasmon frequency to the frequency of light with the same wavevector, $\alpha = \sqrt{e^2n_ew/2\pi\varepsilon_0\m c^2}$.
In our Hall bar, we estimate $\alpha \simeq 0.15$ and thus do not expect significant modification of the MPR dispersion.
It is known, however, that even when retardation effects are not important the actual plasmon frequency is expected to be about 15\% lower than $\op(q_0)$ estimated from \req{eq.p}.\citep{mikhailov:2004,mikhailov:2005}
Such a reduction was observed in both early \citep{vasiliadou:1993} and more recent \citep{kukushkin:2006b,yuan:2006} experiments.

As shown in \rfig{fig1}(b), at $f = 150$ GHz the MPR peak is moved towards the second MIRO maximum and is no longer overlapping with the ZRS.
This observation is consistent with the MPR dispersion relation, \req{eq.mp}, which dictates stronger, compared to MIRO, dependence of the MPR peak position on the microwave frequency.  
Direct comparison of the data at $f=170$ GHz and at $f=150$ GHz reveals roughly equal MIRO amplitudes, indicative of comparable effective microwave intensities incident on our 2DES.
We fit our data (cf. dashed curve) in the vicinity of the MPR peak with $\rho_\omega(B) = a + b/[(B-B_0)^2+(\delta B)^2] + c(B-B_0)$, where the last term accounts for a $B$-dependent background.
Unlike the $f=170$ GHz data, the fit reveals positive background resistance, $a \approx 0.15$ $\Omega$, explaining a considerably higher ($\approx 2.35$ $\Omega$ if measured from zero) MPR peak compared to one at $f=150$ GHz.

\begin{figure}[t]
\includegraphics{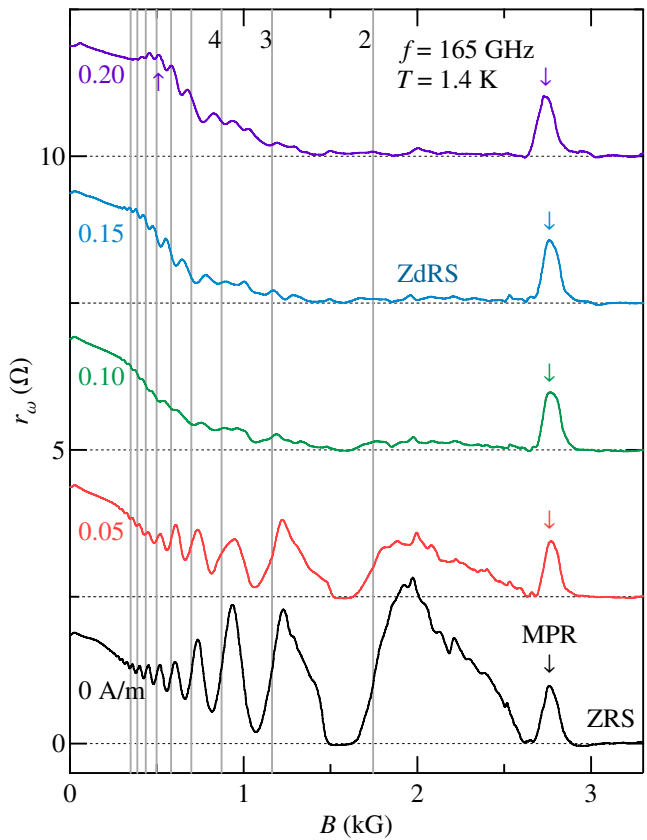}
\vspace{-0.1 in}
\caption{(Color online)
Differential resistivity $r_\omega$ versus magnetic field $B$ under microwave irradiation of $f=165$ GHz measured at $T=1.4$ K for different current densities from $j = 0$ to 0.20 A/m, in a step of 0.05 A/m.
The traces are vertically offset for clarity by 2.5 $\Omega$.
The vertical lines are drawn at the harmonics of the cyclotron resonance, $\omega/\oc=1,2,3,\dots\,$.
}
\vspace{-0.15 in}
\label{fig2}
\end{figure}
We now turn to the role of a dc electric field on MIRO and, especially, on the MPR peak.
In \rfig{fig2} we present the differential resistivity $r_\omega$ as a function of $B$ measured at $T=1.4$ K, under microwave irradiation of $f=165$ GHz, and selected direct current densities from $j\equiv I/w=0$ to 0.20 A/m, in a step of 0.05 A/m.
The traces are vertically offset for clarity by 2.5 $\Omega$ and the vertical lines are drawn at the harmonics of the cyclotron resonance, $\omega/\oc=1,2,3,\dots\,$. 
First, we observe that the response of MIRO to the dc field is strongly nonlinear.
Indeed, already at $j = 0.05$ A/m ($I=2.5$ $\mu$A) MIRO decrease in amplitude by about a factor of two and at $j = 0.10$ A/m almost disappear.
At higher $j$, high-order MIRO reappear and start shifting towards higher $B$, in agreement with previous experimental\citep{zhang:2007c,hatke:2008a,hatke:2011c} and theoretical\citep{lei:2007b,khodas:2008} studies.
This behavior is a result of nonlinear mixing of MIRO and Hall field-induced resistance oscillations,\citep{yang:2002,zhang:2007a,hatke:2009c,hatke:2011a} arising due to electron backscattering off short range disorder between Hall field tilted Landau levels.
On the other hand, the low order MIRO maxima remain strongly suppressed and the data reveal a state with close-to-zero differential resistance which spans a wide magnetic field range.
The presence of microwave irradiation in formation of these states is not essential since they also emerge in a non-irradiated 2DES in a similar range of electric and magnetic fields.\citep{hatke:2010a}

Unlike MIRO, which change dramatically with increasing dc field, the MPR peak shows surprisingly little variation both in magnitude and in position.
This behavior is totally unexpected since it demonstrates that the microwave radiation protects Ohmic behavior within a narrow magnetic field range corresponding to the MPR.
We note that in all previous experiments within this range of dc fields, both with and without microwave radiation, the Ohmic regime always remained limited to an order of magnitude lower magnetic fields.\citep{zhang:2007a,zhang:2007c,hatke:2010a}  

In summary, we have studied microwave photoresistance of a Hall bar-shaped, high mobility GaAs/AlGaAs quantum well. 
In addition to microwave-induced resistance oscillations and zero-resistance states, the photoresistance reveals a distinct peak which originates from a fundamental mode of a dimensional magnetoplasmon resonance.
This MPR peak is several times higher than the dark resistivity, likely due to strongly temperature dependent dark resistivity\citep{hatke:2012a} in our 2DES.
Analysis of the MPR peak, when it is superimposed onto a ZRS, allows us to obtain information about the ZRS-associated absolute negative resistance, which is otherwise masked by instabilities.\citep{andreev:2003}
When the irradiated system is further subject to a dc electric field, microwave-induced resistance oscillations quickly decay and the 2DES exhibits a state with close-to-zero differential resistance. 
The MPR peak, on the other hand, is found to be immune to the dc field exhibiting Ohmic behavior.

We thank J. Jaroszynski, J. Krzystek, G. Jones, T. Murphy, and D. Smirnov for technical assistance.
This work was supported by the US Department of Energy, Office of Basic Energy Sciences, under Grant Nos. DE-SC002567 (Minnesota) and DE-SC0006671 (Purdue). 
A portion of this work was performed at the National High Magnetic Field Laboratory (NHMFL), which is supported by NSF Cooperative Agreement No. DMR-0654118, by the State of Florida, and by the DOE.


\begin{thebibliography}{58}
\bibitem[{not({\natexlab{a}})}]{note:5}
\bibinfo{note}{In our 2DES, $\bar\varepsilon \approx 6.9$, which is the average
  of dielectric constant of GaAs (12.8) and that of free space (1).}
\bibitem[{not({\natexlab{a}})}]{note:x}
\bibinfo{note}{Measured at $B<0$, negative magnetoresistance, MIRO, and MPR photoresistance peak are all of approximately the same strength as their counterparts at $B >0$.}
\end{thebibliography}
\end{document}